\documentclass[amsmath,amssymb,superscriptaddress,twocolumn]{revtex4}
\usepackage[pdftex]{graphicx}
\usepackage{upgreek}
\newcommand{\whz}{$\sqrt{\textrm{Hz}}\,\,\,$}
\usepackage{multirow}

\begin{document}
\title{Optomechanical reference accelerometer}
\author{Oliver Gerberding}
\author{Felipe Guzm\'an Cervantes}\thanks{\scriptsize Corresponding authors: felipe.guzman@nist.gov, jmtaylor@umd.edu}
\affiliation{\footnotesize National Institute of Standards and Technology, 100 Bureau Drive, Gaithersburg, MD\,20899, USA.}
\affiliation{\footnotesize Joint Quantum Institute, University of Maryland, College Park, MD\,20742, USA.}
\author{John Melcher}
\author{Jon R. Pratt}
\affiliation{\footnotesize National Institute of Standards and Technology, 100 Bureau Drive, Gaithersburg, MD\,20899, USA.}
\author{Jacob\,M. Taylor}\thanks{\scriptsize Corresponding authors: felipe.guzman@nist.gov, jmtaylor@umd.edu}
\affiliation{\footnotesize National Institute of Standards and Technology, 100 Bureau Drive, Gaithersburg, MD\,20899, USA.}
\affiliation{\footnotesize Joint Quantum Institute, University of Maryland, College Park, MD\,20742, USA.}
\affiliation{\footnotesize Joint Center for Quantum Information and Computer Science, University of Maryland, College Park, Maryland 20742, USA}
%

\begin{abstract}
We present an optomechanical accelerometer with high dynamic range, high bandwidth and readout noise levels below 8\,$\upmu$g/\whz. The straightforward assembly and low cost of our device make it a prime candidate for on-site reference calibrations and autonomous navigation. We present experimental data taken with a vacuum sealed, portable prototype and deduce the achieved bias stability and scale factor accuracy. Additionally, we present a comprehensive model of the device physics that we use to analyze the fundamental noise sources and accuracy limitations of such devices.
\end{abstract}
\maketitle
%

\section{Introduction}

Accelerometers and gyroscopes form the fundamental building blocks of inertial sensing \cite{Barbour2001,Barbour2010}. Devices with a wide range of bandwidth, precision, accuracy and dynamic range are available and deployed in various applications, including commercial products, medical devices \cite{Cooper2009}, construction engineering \cite{Strasberg1996}, natural resource exploration \cite{Nakstad2008}, inertial sensing for autonomous navigation \cite{Dong2010} and fundamental research \cite{Lenoir2011a,Ignatiev2015}. The critical parameters of an accelerometer with a given bandwidth are its precision, often denoted as acceleration noise $\widetilde{a}_{\rm acc}$; its bias stability, the long-term drift of the DC acceleration $a_{\rm bs}$; and the scale factor $\mathcal{S}$ uncertainty, which describes errors in the value relating physical observable quantities, such as the output voltage $v_{\rm acc}$ to acceleration ($ v_{\rm acc} = \mathcal{S} \cdot {a}_{\rm ext} $, $(\mathcal{S}) = $ 1 V/m/s$^2$ ).

Commercially available accelerometers require laborious calibrations to ensure a minimal deviation of the above described parameters. Such calibrations are often performed at National Metrology Institutes (NMIs) \cite{Robinson1987}, where the test unit is mounted onto an interferometrically-interrogated reference shaker system \cite{Ripper2009,Martens2004,Martens2014}. Such calibrations reach relative uncertainties of the order of $10^{-2} - 10^{-2.5}$. Already calibrated devices, known as reference accelerometers, can, in turn, be used to calibrate further devices using more simple back-to-back measurement set-ups. However, such dependent calibrations are accompanied by an inevitable degradation in uncertainty.

Optomechanical accelerometers interrogated using fiber interferometric methods have recently demonstrated high levels of readout precision \cite{Krause2012,Zhang2013,Guzman2014}, using optical instead of electro-static readout. The devices investigated by Guzman et\,al.\,\cite{Guzman2014} use monolithic fused silica in-plane oscillators interrogated by fiber optic micro-cavities \cite{Rugar1988,Smith2009}. 
These devices have achieved levels of acceleration measurement sensitivities below 100\,ng$/\sqrt{\textrm{Hz}}$ and, more importantly in this context, they provide direct traceability to SI units, a property we denote as ``self-calibrating''. This is achieved in a two-step process: First, external accelerations of the oscillator are converted to displacement via its transfer function, characterized by its resonance frequency $\omega_0$ and mechanical quality factor $Q$. Second, displacement is turned into a readout voltage via laser interferometry. Similar devices have also recently been used as optomechanical force sensors for atomic force microscopy (AFM) \cite{Melcher2014}.

In this article, we present a similar device to Ref.\,\cite{Guzman2014} that is interrogated by a low-finesse cavity formed by a gap of approx. 50\,$\upmu$m between two flat-cleaved fibers \cite{Rugar1988}. Its simple and cost-effective construction, combined with self-calibration, high dynamic range, and high bandwidth (exceeding 10\,kHz) make this device a promising candidate to perform autonomous navigation and on-site acceleration calibrations of other accelerometers. We demonstrate the portability and accuracy of such a device by vacuum packaging it and by characterizing its readout noise and bias stability. Without additional feedback, the low finesse readout allows our device to have a scale factor uncertainty lower than $10^{-3}$ for accelerations from mg to those exceeding 2\,g.  We also present our current understanding of the device physics, optics and mechanics and discuss the expected limits for acceleration readout noise, accuracy and self-calibration and, in addition, we extrapolate this analysis to a high-finesse cavity readout, paving the path for substantial improvements.

\section{Accelerometer prototype}

\begin{figure}
	\centering
	\includegraphics[width=\columnwidth]{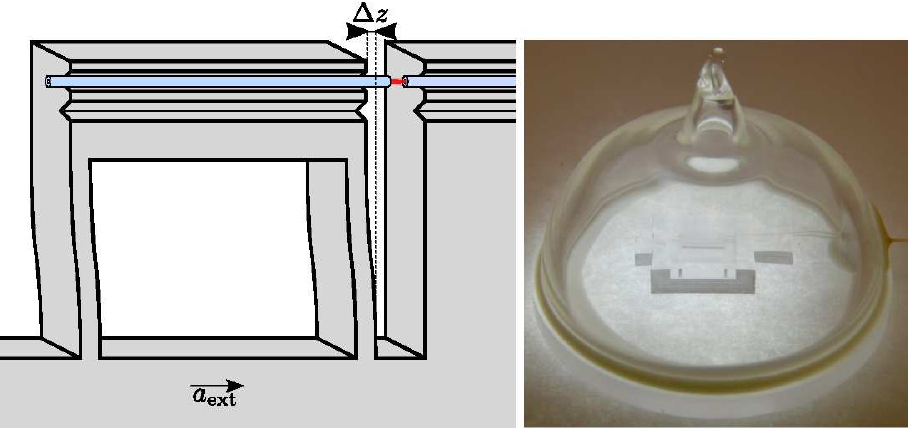}
	\caption{\label{fig:fibercavacc_lowfiness4} (a): Sketch of the acceleration sensing using the mechanical oscillator and a single low-finesse fiber micro-cavity. (b): Photograph of the vacuum sealed accelerometer prototype with two fiber interferometers, one mounted in the lower v-grooves for redundancy.}
\end{figure}
Our accelerometer device (shown in Figure \ref{fig:fibercavacc_lowfiness4}) uses a similar in-plane, monolithic fused-silica oscillator design described by Guzman et.\,al.\,\cite{Guzman2014}. It is frit-bonded onto an additional micro-machined fused-silica part that contains a relief for enabling free test mass motion. This part is, in turn, frit-bonded onto a 4\,cm radius quartz plate with a thickness of 3\,mm. The accelerometer is enclosed by a glass bell, which is glued with Torr Seal onto the base plate. Two flat-cleaved fibers are fed into the bell at a cut-out and their ends are glued into v-grooves next to the test mass.  Flat-cleaved fibers glued onto opposite facing v-grooves on the test mass act as cavity end mirrors.
The acceleration-driven test mass motion translates to optical phase changes of the fiber micro-cavity, which is read out in reflection by monitoring the reflected power.
The fiber ends acting as cavity mirrors, have a low reflectivity of approximately 4\,\%. The oscillator was designed to have a mass of 25\,mg and a resonance frequency of about 10\,kHz. 

\begin{figure}
	\centering
	\includegraphics[width=\columnwidth]{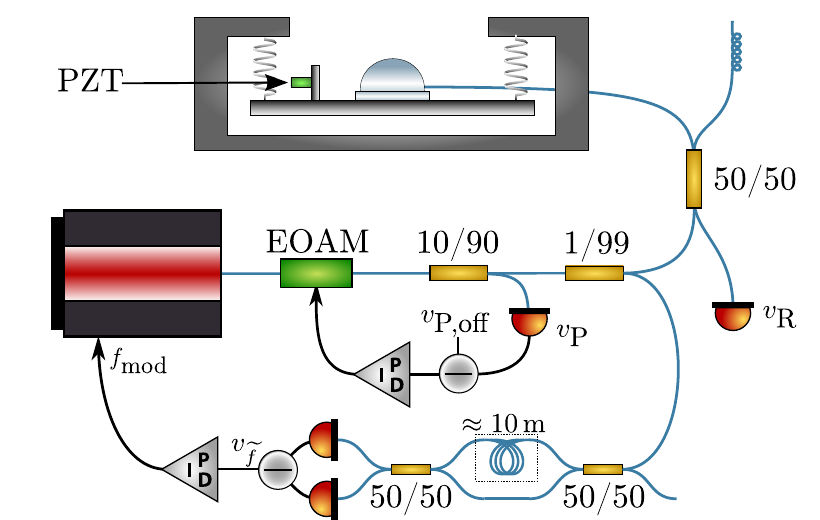}
	\caption{\label{fig:set-up} Experimental set-up for the accelerometer characterization. A laser beam from a widely tunable laser \cite{Newport2014} is send into one of the fiber cavities and the reflected power is detected on a photo receiver ($v_R$). To increase optical power for the performance test an erbium-doped fiber amplifier \cite{Optilab2010} (not shown) was integrated after the laser source. Part of the laser light is split off to stabilize the laser power by actuating on a fiber-based electro-optic amplitude modulator (EOAM). A fiber-based unequal arm-length Mach-Zehnder interferometer, with 10\,m fiber delay between the arms, is used as frequency reference. It is read out via a balanced detection scheme and the error signal is used to stabilize the laser frequency by actuating on the laser pump current and on the laser cavity piezo. The accelerometer is placed on an isolation platform to reduce coupling of unwanted accelerations. A piezo crystal (PZT) mounted on the platform is used to excite the accelerometer. The second, unused output port of the 50/50 beam splitter was additionally terminated by wrapping the fiber with a very small bending radius and thereby introducing excess losses of the light traveling through the fiber core.}
\end{figure}

After assembling the fiber cavities, we vacuum sealed our device by pumping on a glass pipe
originally connected to the top of the bell and performing a vacuum pulling by flame heating the pipe. This is evident by the residual glass structure on top of the bell shown in Figure \ref{fig:fibercavacc_lowfiness4}b. The device is now available as a portable accelerometer which we connected to our measurement set-up shown in Figure \ref{fig:set-up}. 

\subsection{Laser wavelength scan}
\label{ssec:lasscan}

\begin{figure}
	\centering
	\includegraphics[width=\columnwidth]{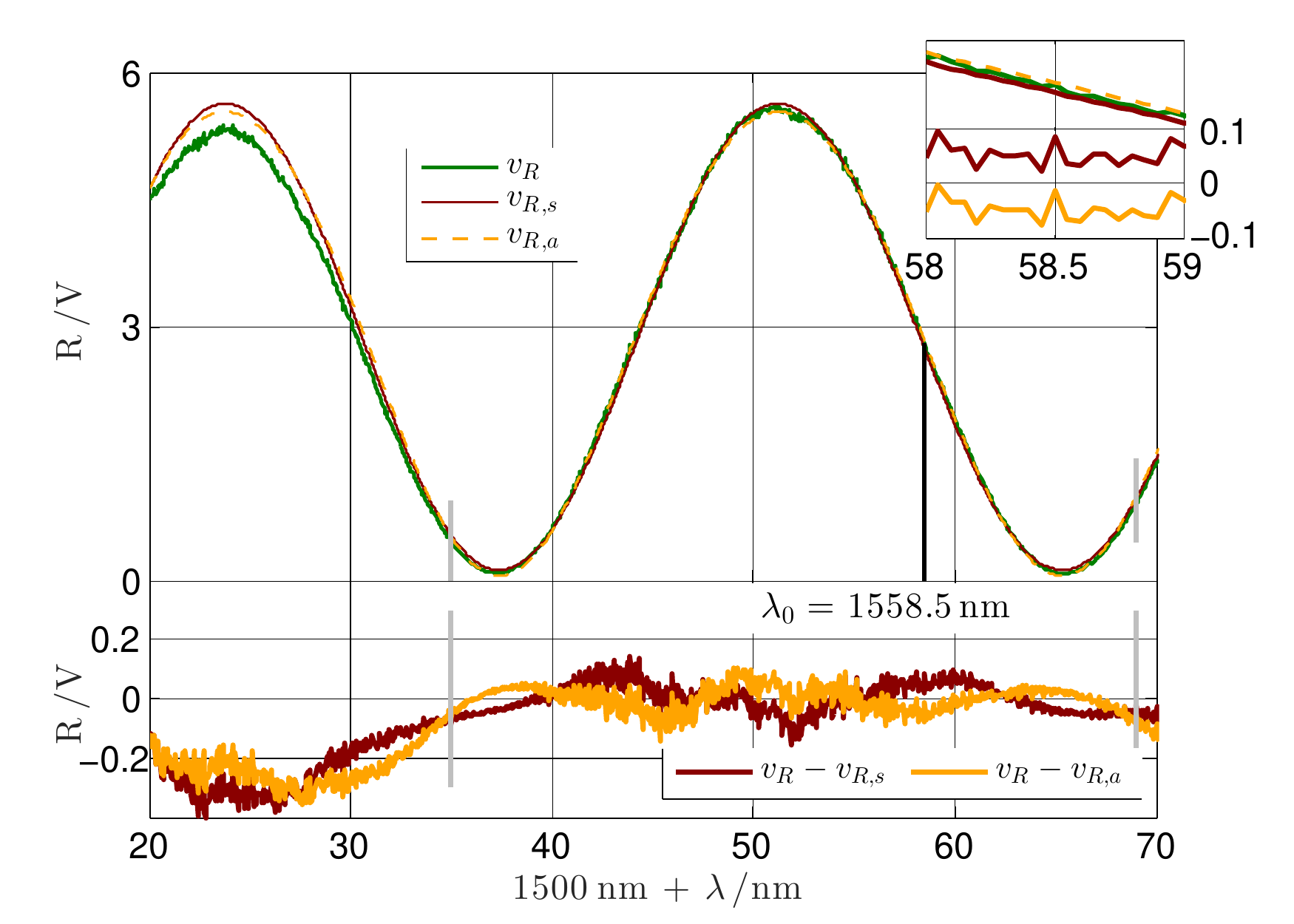}
	\caption{\label{fig:laser_scan_model} The top shows the measured cavity response $v_{R}$ to a wavelength sweep in comparison to the fitted sinusoidal ($v_{R,s}$) and Airy function ($v_{R,a}$) models. The bottom plot shows the residuals of the two fits. The gray, vertical lines indicate the range of data used for the fits. The laser power was actively stabilized during the sweep to account for a changing laser output power. The inset in the top right corner shows a zoom into the quadrature point for the response and the residuals.}
\end{figure}
We use a widely tunable laser \cite{Newport2014} to measure the dependency of the reflected light power $P_R$, and the corresponding photo detector output voltage $v_R$ at the laser wavelength $\lambda$. Figure \ref{fig:laser_scan_model} shows the response for one of the two cavities, which was used throughout this study.

We first fit our data using the following sinusoidal model approximation of the fiber cavity response.
\begin{equation}
v_{R,s} =  v_{\mathrm{DC}} - v_{\mathrm{DC}} \cdot \kappa \cos ( 4 \pi \frac{z_m}{\lambda} ) ,
\label{eq:cavrespsinevoltagesimple}
\end{equation}
where $v_{\mathrm{DC}}$ is the DC output voltage measured in reflection, $\kappa$ is the optical contrast, and $z_m$ is the macroscopic cavity length. By fitting our measurement data with this response we determine $v_{\mathrm{DC}} = 2.85$\,V, $\kappa = 0.97$ and $z_m \approx 42.6\,\upmu m$. We found a low uncertainty of about 0.15\,\% for the slope at the critical quadrature points of the response between the fit and the data. The residuals indicate that our response function models are not able to fully describe the measured behavior. We attribute this to effects related to diffraction, residual misalignments and coupling efficiency into the fiber mode \cite{Wilkinson2011}, which become relevant due to our short, non mode-matched cavity and the wide tuning of  laser wavelength. However, the impact of these effects is much lower around our operating point. This is discussed some more in Section \ref{sec_accuracy}.  By tuning the laser wavelength to a quadrature point of the response ($\lambda_q$), we maximize the ratio between voltage and length change to a now calibrated value \cite{Rugar1989}.
%
\begin{equation}
\Delta v_{R} =  v_{\mathrm{DC}} \cdot \kappa   \frac{4 \pi}{\lambda_q} \Delta z_m .
\label{eq:cavrespsinevoltagesimplequad}
\end{equation}

We also analyze our measurement data with the Airy-function, which fully takes multiple reflections into account \cite{Smith2009}:
\begin{equation}
v_{R,a} =  v_{\mathrm{off}} + v_{\mathrm{\gamma}}  \frac{(1+R)^2}{2} \left[  \frac{1-\cos\left( \frac{4 \pi}{\lambda} z_m\right)}{ 1 + R^2 - 2R \cos \left( \frac{4\pi}{\lambda} z_m\right)} \right].
\label{eq:airyfunc}
\end{equation}
Here we define an offset voltage $v_{\mathrm{off}}$, an amplitude scaling voltage $v_{\mathrm{\gamma}}$ that includes the optical contrast and the reflectivity $R \approx 0.04$. Around the now slightly shifted point of maximum slope, $\lambda_{0}$, we can again define an effective linear coupling of displacement into voltage in a more general form,
\begin{equation}
\Delta v_{R,a} =  \frac{\lambda_{0}}{\bar{z}_m} \left( \frac{\mbox{d}
	 v}{\mbox{d} \lambda} \right) \Delta z_m.
\label{eq:airyfunceff}
\end{equation}
%
\subsection{Ring down}

\begin{figure}
	\centering
	\includegraphics[width=\columnwidth]{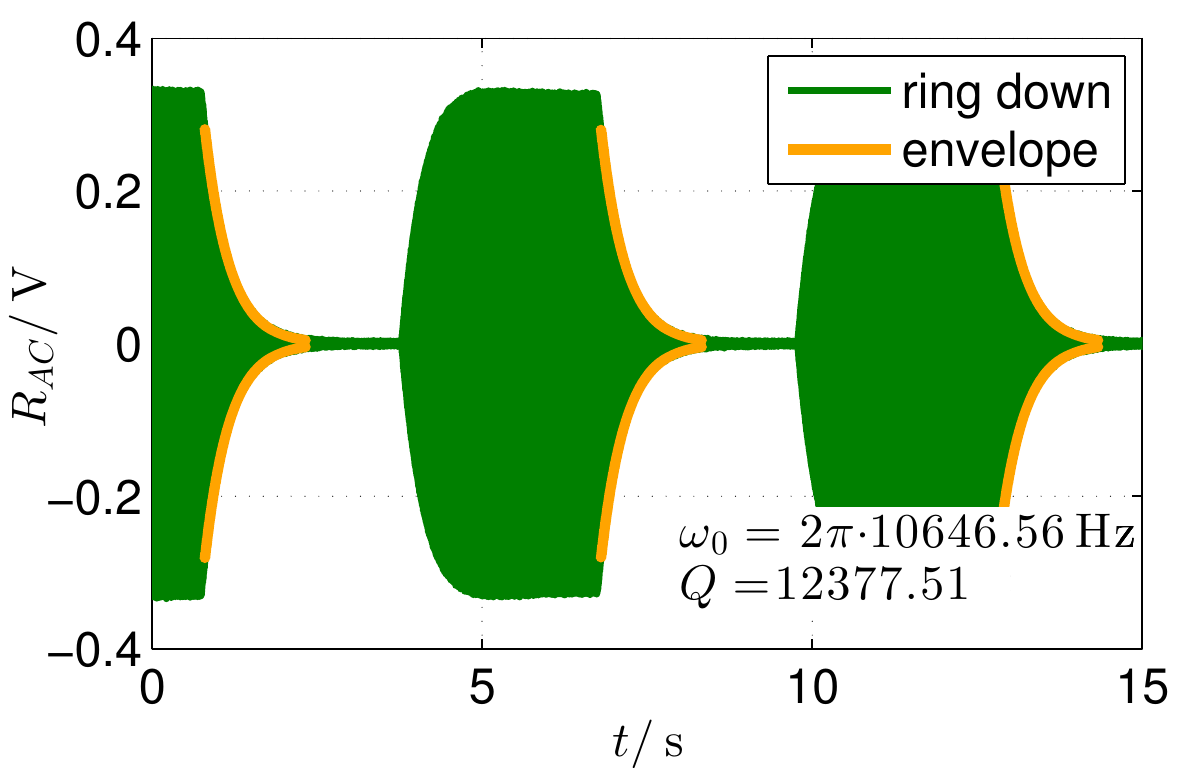}
	\caption{\label{fig:RD_TS} Oscillator ring downs in comparison to the fitted envelope functions. The measured data was high-pass filtered to exclude low frequency noise and to make the decay clearly visible.}
\end{figure}
To estimate the oscillator mechanical characteristic parameters we perform a ring-down experiment with our device. We excite it close to the resonance frequency, turn off the excitation, and then monitor the oscillation decay. A measurement for our device is shown in Figure \ref{fig:RD_TS}. We fit our data using the formula
\begin{equation}
{v}_{RD} =  {v}_{0} \cdot e^{- \frac{\omega_0 t}{2 Q}} \sin{(\omega_0 t + \phi)}.
\label{eq:ringdownamp}
\end{equation}
From this fit we estimate the natural frequency of the oscillator $\omega_0$ ($\approx 2\pi\times 10646.56$\,Hz) with a fit standard error on the order of $10^{-8}$. We then perform a software demodulation of the signal at $\omega_0$ to extract only the exponential decay of the oscillation amplitude $\bar{v}_{rd}$, which is governed by the following equation.
%
\begin{equation}
\bar{v}_{rd} =  \bar{v}_0 \cdot e^{- \frac{\omega_0 t}{2 Q}}.
\label{eq:ringdownamp}
\end{equation}
By fitting this formula to the decay we determine the quality factor $Q \approx 12379.07$ with a standard uncertainty of about $15 \times 10^{-6}$. The resulting fit curve is shown in Figure \ref{fig:RD_TS}, as the envelope of the ring down. A correction of $\omega_0$ due to a damping-induced frequency shift is not necessary, since the correction factor is fully negligible at these levels of $Q$. To further refine and test the long term stability of $\omega_0$ and $Q$ we perform the above described analysis for multiple, consecutive ring down measurements. For 250 ring downs, spaced over 25\,minutes (1500 seconds) of measurement time, we determine a standard deviation for $\omega_0$ on the order of $0.1 \times 10^{-6}$ and $200 \times 10^{-6}$ for $Q$, each an order of magnitude larger than the single-shot statistical variation. Using the transfer function of a damped harmonic oscillator, we have the relation between external accelerations and displacement:
\begin{equation}
T_{\mathrm{HO}} (\omega) = \frac{\Delta z(\omega)}{a_{\mathrm{ext}}(\omega)} = - \frac{1}{\omega_0^2 - \omega^2 + i \frac{\omega_o}{Q}\omega}.
\label{eq:TFaccdisp1}
\end{equation}

\subsection{Noise performance}

\begin{figure}
	\centering
	\includegraphics[width=.95\columnwidth]{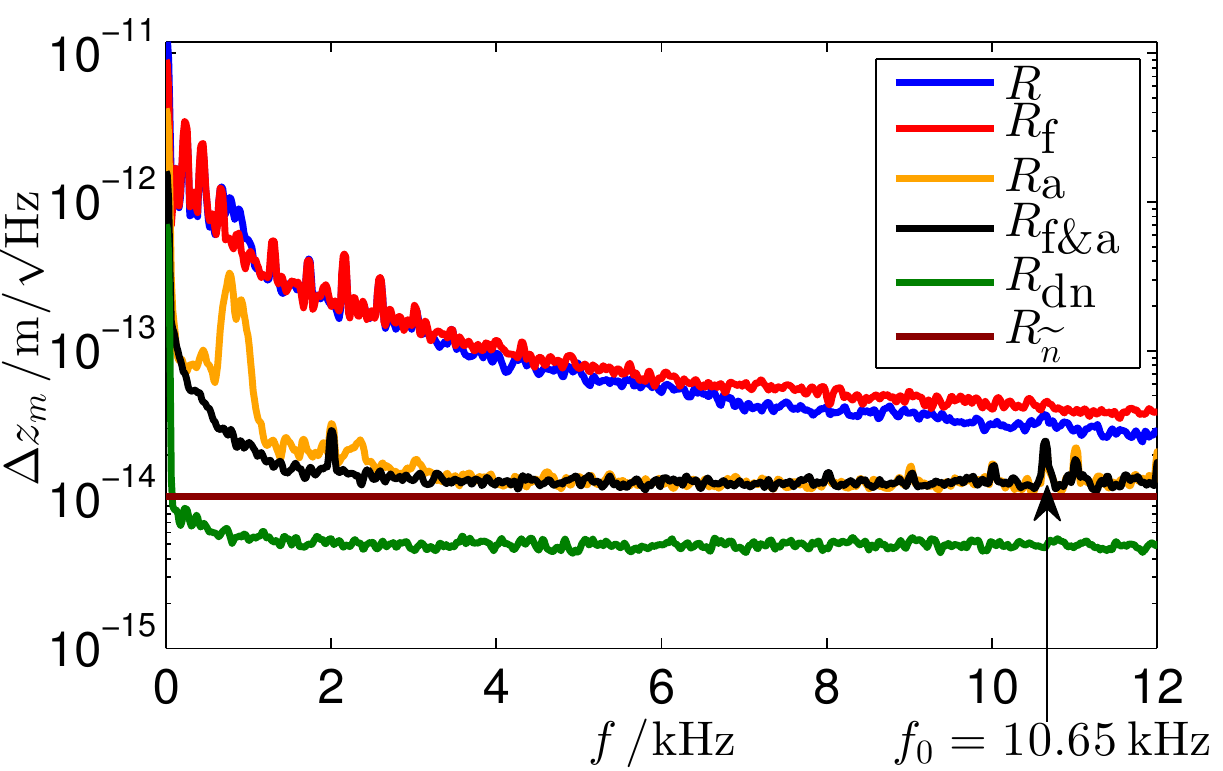}
	\caption{\label{fig:spec_comparison_length} Measured displacement spectra without stabilizations $(R_{})$, with amplitude stabilization $(R_{\textrm{a}})$, frequency stabilization $(R_{\textrm{f}})$ and both stabilizations $(R_{\textrm{f\&a}})$ running. Shown are also the estimated shot noise level $(R_{\widetilde{n}})$ and the measured dark noise $(R_{\textrm{dn}})$. The thermally excited resonance peak of the mechanical oscillator is visible at $f_0$.}
\end{figure}

With our device resting on a vibration isolation platform we measure the spectra of the output voltage to determine the readout noise floor. Using equations \ref{eq:cavrespsinevoltagesimplequad} and \ref{eq:TFaccdisp1} we convert the voltage spectra into the corresponding displacement. The results of this are shown in Figure \ref{fig:spec_comparison_length} for different laser stabilization schemes (see Figure \ref{fig:set-up}). With both laser amplitude and frequency stabilization the spectrum shows the thermally excited peak of the oscillator resonance, which sticks out of an almost flat noise floor of about 15\,fm/\whz. The noise increases at lower frequencies, though the magnitude is strongly reduced by the stabilizations. We estimate the shot noise as explained in Section \ref{sssec:shot} with a measured DC power on the photodiode of 60\,$\upmu$W. 
The dark noise measurement includes contributions from the photo receiver and the measurement devices. This was measured by blocking all light going onto the measurement photodiode. The laser frequency stabilization allowed us to strongly reduce the excess frequency noise, likely induced by coupling of acoustics and vibrations into the tunable laser head \cite{Harvey1991,Hawthorn2001}. The unequal armlength Mach-Zehnder configuration is able to operate at any given wavelength and can, therefore, be used with the tunable laser, necessary to perform the self-calibration.

\begin{figure}
	\centering
	\includegraphics[width=0.95\columnwidth]{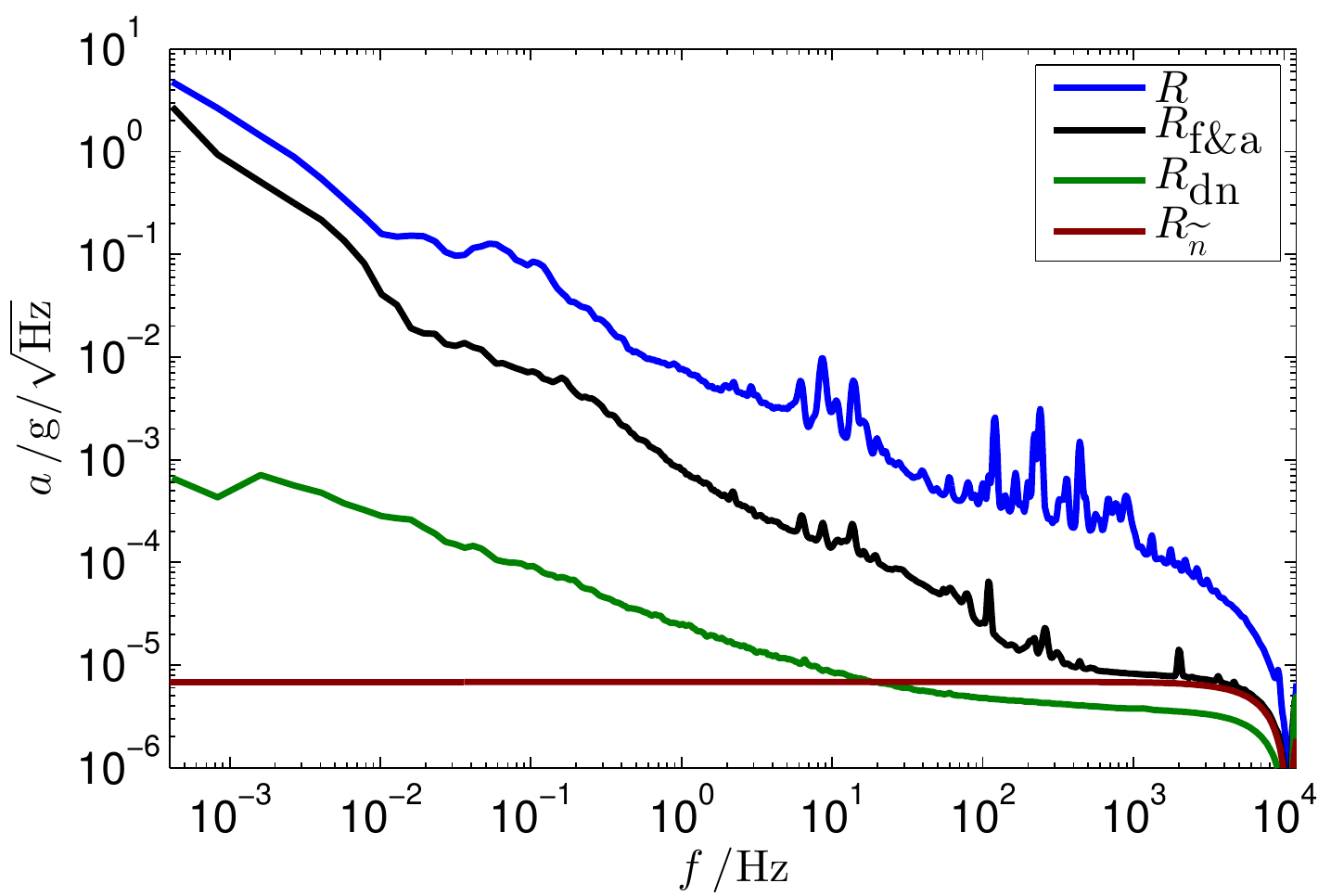}
	\caption{\label{fig:spec_comparison_acceleration_g_log} Measured acceleration spectra with $(R_{\textrm{f\&a}})$ and without stabilizations $(R_{})$ running. Shown are also the estimated shot noise level $(R_{\widetilde{n}})$ and the measured dark noise $(R_{\textrm{dn}})$.}
\end{figure}

Using the harmonic oscillator transfer function model of the mechanical oscillator we can determine the corresponding acceleration noise spectra for a given displacement noise. Using a data acquisition system we performed long-term measurements. We converted the measured voltage into acceleration by filtering and scaling it accordingly. The resulting acceleration noise spectra are shown in Figure \ref{fig:spec_comparison_acceleration_g_log}. 
Our device achieves an acceleration noise floor better than 8$\upmu$g/\whz above 1\,kHz. Its low frequency performance is limited by a $1/f$ noise, leading to levels of 900\,$\upmu$g/$\sqrt{\textrm{Hz}}$ at 1\,Hz and better than 40\,mg/$\sqrt{\textrm{Hz}}$ at 10\,mHz (corresponding to a displacement noise of 2\,pm/$\sqrt{\textrm{Hz}}$ and better than 90\,pm/$\sqrt{\textrm{Hz}}$ respectively).

\subsection{Allan variance}
\begin{figure}
	\centering
	\includegraphics[width=\columnwidth]{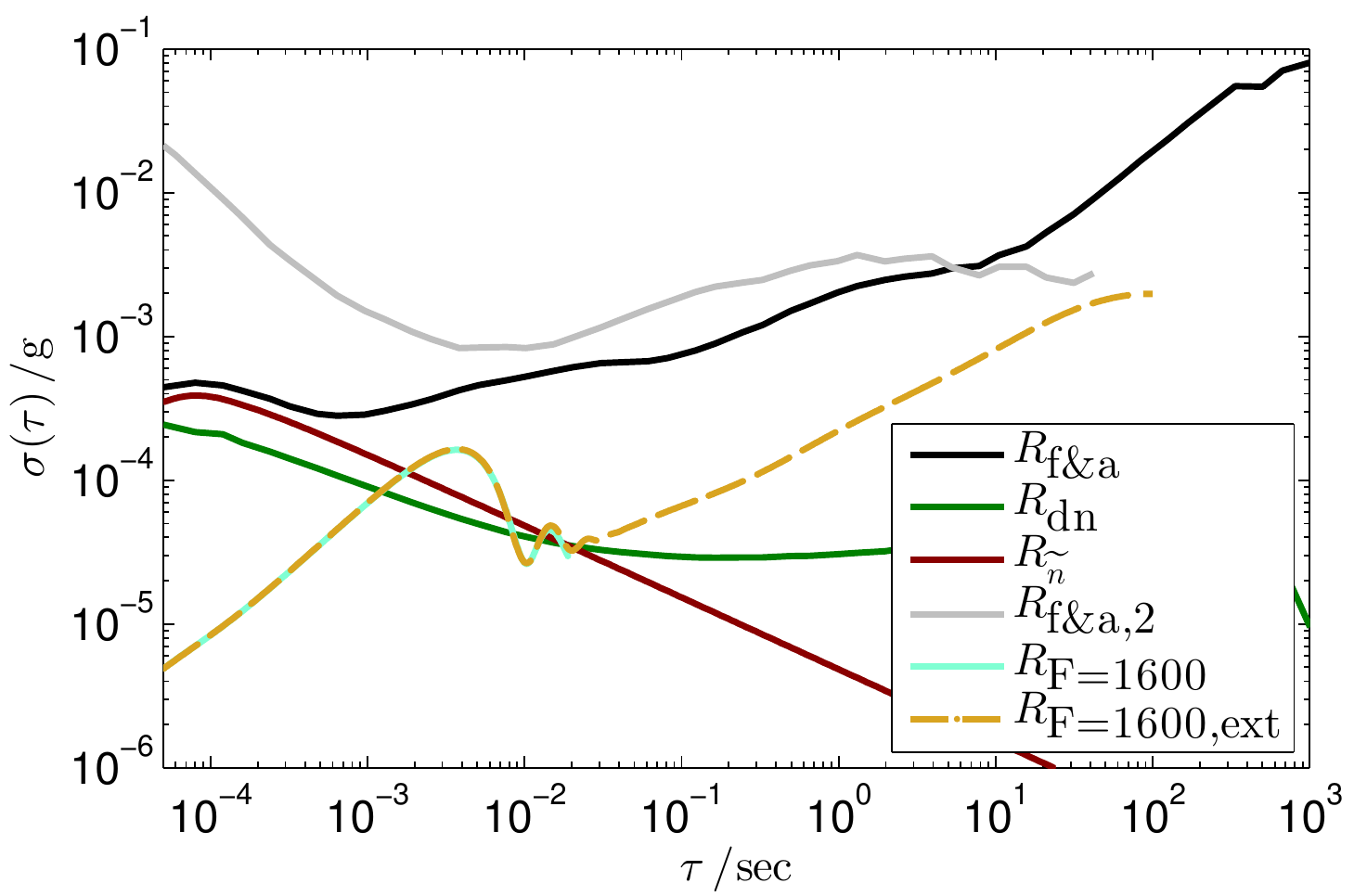}
	\caption{\label{fig:allan_dev_acc_g} Measured Allan deviations of our detector output signal converted into acceleration. Shown are the results with both stabilizations $(R_{\textrm{f\&a}})$, the estimated shot noise level $(R_{\widetilde{n}})$ and the measured dark noise $(R_{\textrm{dn}})$. A second measurement with both stabilizations and a higher readout dark noise floor $(R_{\textrm{f\&a,2}})$ is shown as well. It shows an improved stability at longer integration times, indicating that the noise is not constant at longer integration times. 
	For comparison we also show the Allan deviations calculated from the measured power spectral densities for the high-finesse cavity readout $(R_{\textrm{F=1600}})$ \cite{Guzman2014}. We also extended the corresponding spectrum with a 1/f noise model expected for laser frequency noise of about 200kHz/$\sqrt{\textrm{Hz}}$ at 1\,Hz $(R_{\textrm{F=1600,ext}})$ and with a cavity length of 100\,$\upmu$m (see Equation \ref{eq:freqnoise_phase}).}
\end{figure}

To investigate the bias stability we determined the Allan deviation of our readout. 
The results are shown in Figure \ref{fig:allan_dev_acc_g} with and without stabilization. At short integration times we are limited only by shot noise. The exact nature of the increase at longer integration times is currently under investigation, and it is presumed to be caused by residual frequency noise or parasitic stray beams in our fiber set-up. Each of these contributions is susceptible to acoustic and thermal fluctuations, causing the noise in our set-up to not be constant.
This is evident from the second, shorter measurement with stabilizations shown in Figure \ref{fig:allan_dev_acc_g}, which presents the best levels of long-term stability achieved during our measurement campaign, even though the readout noise floor was significantly higher.
Our frequency reference Mach-Zehnder interferometer was placed in the same thermal environment as our laser source and was isolated passively against thermal fluctuations. The use of active thermal stabilizations for the fiber interferometer may help to improve the performance in future implementations. We reach the minimum of our Allan deviation at about 1\,ms of integration time with a value of $3\times10^{-4}\,$g. 

During our investigation we found that the use of fiber circulators, to feed light to the accelerometer and to detect the reflection, caused an increased coupling of laser frequency noise into the measurement. We attribute this to parasitic beams, introduced by the excess leakage in the return path of the circulator, that are phase modulated relative to our signal of interest by the frequency noise. These beams, or stray light, can reach significant amplitudes in comparison to the reflected signal of the low finesse cavity, which is only about 4\% of the incident power, and contaminate the interfered signal, leading to enhanced coupling of laser frequency noise and other phase noise in our fiber set-up. Using a 50/50 fiber beam splitter we were able to actively tune and ultimately minimize this coupling by terminating the open, unused output of the splitter. This improvement in stability comes at the cost of reducing the detected power to about one quarter of the levels achieved with a circulator. Future implementations have to take these parastic signals into account to reduce the coupling of laser frequency noise to the signal from the actual cavity. 

As an example of what may be possible with such a device having a higher-finesse optical readout,
Figure \ref{fig:allan_dev_acc_g} shows Allan deviations for the high finesse cavity readout implemented by Guzman et al. \cite{Guzman2014}, which were measured with a finesse of 1600 and a fixed-frequency laser. The extended curve shows levels corresponding to an extension of the measured spectral density down to 1\,mHz (data was available down to 20\,Hz) with a readout dominated by an assumed laser frequency noise with 191\,kHz/$\sqrt{\textrm{Hz}}$ at 1\,Hz, levels that are regularly achieved with rather simple laser stabilization schemes, and a cavity length of 172\,$\upmu$m. The strongly reduced readout noise floor for this system leads to Allan deviations below $10^{-5}\,$g.

\section{Metrological analysis}

Any external acceleration measured with our devices is converted into a readout voltage. In the following, we describe and model this conversion and its self-calibration as a chain of individual transducer steps. This allows us to i) determine readout noise of the devices, scale factor uncertainty, bias stability and dynamic range for a given set of parameters, and ii) to determine the parameters that are necessary to reach a desired measurement uncertainty.

\subsection{Mechanical oscillator}

\begin{figure}
	\centering
	\includegraphics[width=6cm]{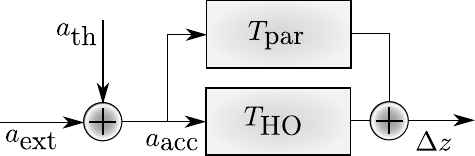}
	\caption{\label{fig:mechmodel} Simplified model of the metrology chain from external oscillator acceleration to test mass displacement. }
\end{figure}

Our devices use an in-plane monolithic oscillator made of fused-silica to convert external accelerations $a_{\mathrm{ext}}$ into measured test mass displacement $\Delta z$ (recall Figure \ref{fig:fibercavacc_lowfiness4}). A block diagram of this conversion is shown in Figure \ref{fig:mechmodel}. The high stiffness and the correspondingly small test mass displacements of our device lets it be described as a damped harmonic oscillator (HO) with an angular resonance frequency $\omega_0$, a mechanical quality factor $Q$ and an effective oscillator mass $m$. The conversion of acceleration into displacement is given, in the frequency domain, by the transfer function $T_{\mathrm{HO}}$ described in Equation \ref{eq:TFaccdisp1}.
Correspondingly, a measured displacement in the time domain is converted back into an acceleration by computing
\begin{equation}
a_{\mathrm{acc}} (t) = -  ( \Delta\ddot{ z}(t) + \frac{\omega_0}{Q} \Delta\dot{ z}(t) + \omega_0^2 {\Delta z}(t) ).
\label{eq:funcaccdisp}
\end{equation}
\subsection{Thermal noise}
The acceleration readout is fundamentally limited by thermal acceleration noise $\widetilde{a}_{\mathrm{th}}$ induced by the finite temperature of the oscillator and mechanical losses \cite{Yasumura2000}.
\begin{equation}
	\widetilde{a}_{\mathrm{th}} = \sqrt{ \frac{4k_{\mathrm{B}} T \omega_0 }{mQ}}
	\label{eq:aacctherm}
\end{equation}
This is approximately a white noise that is simply added to $a_{\mathrm{ext}}$, and therefore, only contributes to the readout noise floor, but it does not influence the scale factor, nor the bias stability.

\subsection{Conversion into displacement}
The scaling of acceleration into displacement depends only on two parameters, $\omega_0$ and $Q$, and on the signal frequency $\omega$, as indicated by Equation \ref{eq:TFaccdisp1}. We can estimate the scaling at two critical points of the frequency response. At DC and at low frequencies the scaling simplifies to $T_{\mathrm{HO}} (\omega \ll \omega_0) = - 1/\omega_0^2$, with no $Q$ dependency. Near resonance the influence of $Q$ is maximal and $T_{\mathrm{HO}} (\omega = \omega_0) = - Q/\omega_0^2$. 

The estimation of both parameters is critical for the scale factor accuracy. The measurement of $\omega_0$ is done by comparing the oscillation to a frequency standard, which, in turn, makes the uncertainty of the standard one of the parameters fundamentally limiting the achievable scale factor uncertainty. The measurement uncertainty of $Q$, which can be determined with different techniques, is only limited by the integration time and the readout noise floor. Larger values of $Q$ require in general a longer measurement time, due to the increase in relaxation time $\tau = \frac{2 Q}{\omega_0}$.

\subsection{Deviations from the harmonic oscillator}
As briefly mentioned above, the harmonic oscillator model is only an approximation of the real behavior of the sensor. A full model that allows us to estimate the error of our approximation is beyond the scope of this article. However, we present the relevant parasitic effects, included as additional transfer function $T_{\mathrm{par}} $ in our model, by separating them into three categories and we discuss their expected behavior. 

\subsection*{Higher order modes}
\begin{figure}
	\centering
		\includegraphics[width = 8cm]{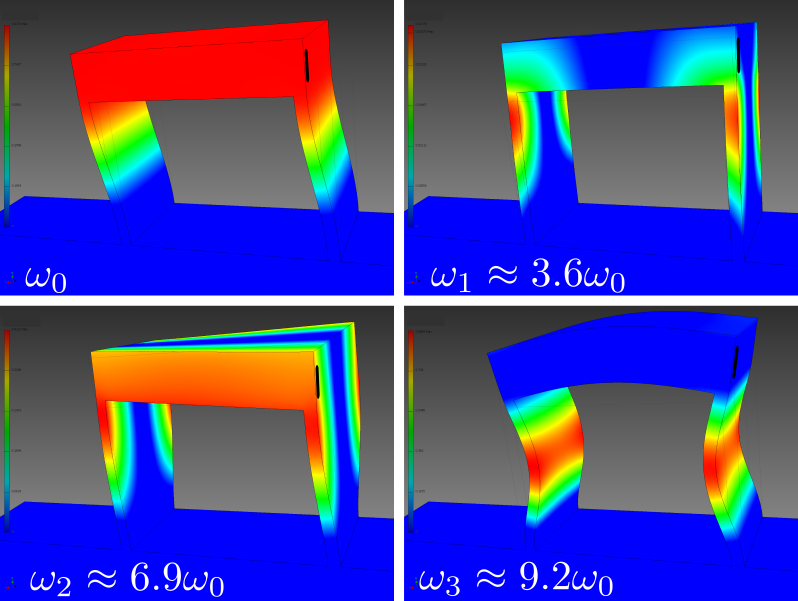}
	\caption{\label{fig:modessimulation} Shown are the ground mode and the three next higher order modes of the oscillator, calculated using Autodesk Inventor finite element analysis. The colour encoding represents the critical displacement in z-direction, the scaling is arbitrary. The detection area is sketched with black stripes. }
\end{figure}

The parallelogram design of our oscillator ensures that higher order modes are well separated in frequency and that their main axis of motion is perpendicular to the fundamental mode. Figure \ref{fig:modessimulation} shows the results of a finite element model analysis of our sensors, which verifies the mode spacing, as well as the reduced coupling of higher order modes into the critical z-axis. As we can see from the plot, the detection of the oscillator motion close to the top surface is not ideal for reducing coupling 
of the higher modes, especially $\omega_2$, into $z$.
Hence, future devices might use larger v-grooves or otherwise optimized geometries to detect the displacement at the optimal point, if these modes prove to be limiting in the future. 

\subsection*{Non-viscous damping}
Our oscillator devices are designed to be operated in vacuum and under small effective displacements, significantly less than 1\,nm. Hence, residual gas damping and mechanical losses inside the oscillator are the dominant contributions, which are well described by a linear viscous damping model. Variations in these parameters are not at all critical, since they are incorporated into the measured $Q$ (providing that these parameters do not change significantly between the measurement time and the self-calibration). Non-viscous damping effects, like structural damping, can result in a different transfer function from acceleration into displacement. One direct way to investigate such effects is to excite the accelerometer with known amplitudes at various frequencies to verify the transfer function, an experiment that will be conducted at a later point. 

\subsection*{Linearity}
Simple beam deflection theory, which is part of the basis for the linear harmonic oscillator model, breaks down for large displacements, at which point additional higher order dependencies on the input frequency become relevant and the fundamental mode loses energy, effectively decreasing $Q$. These effects limit the dynamic range of the oscillator, which relates, for a given readout uncertainty, to a maximum acceleration. One can characterize these behaviors experimentally, by operating the device under the desired maximum acceleration. For low frequencies such a measurement has to determine whether higher harmonics of the excitation frequency are present in the readout. For frequencies close to the resonance this can be combined with a measurement of $Q$ under varying levels of excitation. The total deflection for accelerations of up to 2\,g is on the order of 5\,nm. This is significantly smaller than the thickness of the flexure in z-direction (255\,$\upmu$m), which leads us to expect this contribution to be very small.

\subsection{Fiber cavity readout}
\label{sssec:cav_resp}
\begin{figure}
	\centering
	\includegraphics[width=7cm]{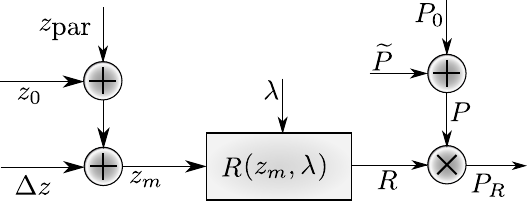}
	\caption{\label{fig:cavmodel} Model of the metrology chain from test mass displacement to reflected power.}
\end{figure}

Our model for the propagation of the test mass displacement $\Delta z$ into reflected optical power $P_R$ is shown in Figure \ref{fig:cavmodel}. 

\subsection*{Cavity length}
The total cavity length is a combination of a constant term $z_0$, the test mass displacement $\Delta z$, and any unwanted, parasitic influences ${z}_{\mathrm{par}}$ caused, for example, by thermal expansion of the sensor. Using the laser wavelength sweep measurement, the total cavity length can be determined in situ. Time dependent parasitic changes will cause an unwanted proportional output signal, inducing noise and decreasing the bias stability, and, for significant changes, a scale factor change. 

\subsection*{Cavity response}

The effective cavity reflectivity for a low-finesse external fiber micro-cavity has been studied in great detail \cite{Wilkinson2011,DiDonato2013,Wilkinson2013}. 
Models are available that include the influence of angular misalignment and multiple reflections. 
In case of low reflectivity one can approximate this function with the sinusoidal response for a simple two beam interference, neglecting all geometric and Gaussian beam effects. The resulting reflected intensity $R$ of the cavity is then given as
\begin{equation}
	R_{\mathrm{}} = R_{\mathrm{DC}} \cdot  \left( 1- \kappa \cdot \cos( 4 \pi  \frac{z_m}{\lambda} )\right) .
	\label{eq:sine}
\end{equation}
Experimentally, this response is implicitly characterized by the laser wavelength scan described in Section \ref{ssec:lasscan}. Since the real response of the cavity is somewhat more complicated, it is expected that this approximation will limit the scale factor accuracy. In the following, we will present a few equations for the sinusoidal model, which are simple and useful for calculating errors due to shot and laser intensity noise for the low-finesse cavity, and we will present the more general formulas for the Airy function model, which is applied to determine the scale factor and for cavities with higher finesse. For the linearized sinusoidal model the reflected intensity around the quadrature point $\lambda_q$ can be estimated 
\begin{equation}
R_{\mathrm{}} = R_{\mathrm{DC}} + \Delta R_{\mathrm{}} =  R_{\mathrm{DC}} + \frac{ R_{\mathrm{DC}} \cdot \kappa \cdot 4 \pi \cdot z_m}{\lambda_q}.
\label{eq:sine2}
\end{equation}
From this approximation it becomes clear that the response depends on the laser wavelength, as well as its stability. Changes in $\lambda$ will move the operating point that was used during measurement characterization, independent of the applied response model, which changes the effective scale factor and bias.

Using the Airy function model (See Equation \ref{eq:airyfunc}), which includes multiple reflections, we generalize the cavity reflected intensity around a given operating wavelength $\lambda_0$ as
\begin{eqnarray}
R_{\mathrm{}} &= R(z_m, \lambda_0) + \frac{\lambda_{0}}{\bar{z}_m} \left( \frac{\mbox{d} R}{\mbox{d} \lambda} \right) \Delta z_m.\\
&= R_{\mathrm{DC}} + \frac{\lambda_{0}}{\bar{z}_m} \left( \frac{\mbox{d} R}{\mbox{d} \lambda} \right) \Delta z_m
\label{eq:genresp}
\end{eqnarray}
In case of a hypothetical perfect cavity model, the laser wavelength uncertainty represents the fundamental limit for the displacement readout accuracy.

\subsection*{Nonlinearity}
For large displacements at a given uncertainty the linear approximations of Equations \ref{eq:sine} and \ref{eq:genresp} introduce errors in the scale factor and generate signals at higher harmonics of the input signal. The dynamic range of our device is relatively large compared to systems using higher values of finesse, and rough error estimates can be easily calculated from the deviations of the linear approximations relative to the expected response functions. We determine a scale factor error of less than $10^{-3}$ for accelerations exceeding 2\,g.  

\subsection*{Laser frequency noise}
Changes in the laser frequency (which are inverse to wavelength fluctuations) couple into the readout as phase noise, causing an effective displacement noise. This is caused by the interference between the reflections, which are delayed relative to each other by twice the cavity length. The effective coupling for small delays in comparison to the readout frequencies is given as
\begin{equation} 
\widetilde{z}_{f} = z_m  \frac{\lambda_0}{c}  \widetilde{f}. 
\label{eq:freqnoise_phase}
\end{equation}
Here $c$ is the speed of light and $\widetilde{f}$ is the laser frequency noise.
\subsubsection{Input power}
The reflectivity of the cavity is sensed with the power sent into the fiber.
\begin{equation} 
P_{R} = P R_{\mathrm{DC}} + P  \Delta R_{\mathrm{}}
\label{eq:cavrespsimplpower}
\end{equation}
We determine the change in reflected power for the Airy function model as
\begin{equation}
\Delta P_{R} =  P \frac{\lambda_{0}}{\bar{z}_m} \left( \frac{\mbox{d} R}{\mbox{d}\lambda} \right) \Delta z_m  = \frac{\lambda_{0}}{\bar{z}_m} \left( \frac{\mbox{d} P}{\mbox{d} \lambda} \right) \Delta z_m.
\label{eq:cavrespsinepower}
\end{equation}
%
\subsection*{Intensity fluctuations}
Relative amplitude/intensity noise (RIN) has to be taken into account as well. Its influence can be described as two separate effects. The first effect is that RIN directly influences the scaling factor, as evident from Equation \ref{eq:cavrespsinepower}. The second effect is a coupling into the readout as additive amplitude noise. To quantify this effect we write the optical power $P$ in terms of a constant component and an additive power noise $\widetilde{P} $,
\begin{equation}
P =   P_0 \cdot \mathrm{RIN} =  P_0 + \widetilde{P} .
\label{eq:RIN}
\end{equation}
The DC component of the optical signal can now be rewritten as a constant term and a fluctuation noise term.
\begin{equation}
 P \cdot R_{\mathrm{DC}} =  P_{0} \cdot R_{\mathrm{DC}} +  \widetilde{P}  \cdot R_{\mathrm{DC}} =  P_{0,\mathrm{DC}} +  \widetilde{P}_{\mathrm{RIN}}.
\label{eq:RINnoise}
\end{equation}
The effective length noise $\widetilde{z}_\mathrm{P}$ due to RIN can be estimated, by computing the ratio of $\widetilde{P}_{\mathrm{RIN}}$ and $\Delta P_{R}/ z_m$. For the sinusoidal model this corresponds to
\begin{equation}
\widetilde{z}_\mathrm{P} =  \frac{\widetilde{P} \lambda_q}{P 4 \pi \kappa} = \mathrm{RIN}\cdot \frac{ \lambda_q}{ 4 \pi \kappa}.
\label{eq:RINnoiselengthsine}
\end{equation}
For the more general cavity response we can write this as
\begin{equation}
	\widetilde{z}_\mathrm{P} =  \frac{\widetilde{P} }{P}   \frac{ R (z_m, \lambda_0) }{ \left( \frac{\mathrm{d} R}{\mathrm{d}\lambda} \right)  } \frac{ \bar{z}_m }{ \lambda_0 } = \mathrm{RIN}\cdot  \frac{ R (z_m, \lambda_0) }{ \left( \frac{\mathrm{d} R}{\mathrm{d} \lambda} \right)  } \frac{ \bar{z}_m }{ \lambda_0 } .
	\label{eq:RINnoiselength}
\end{equation}
Here we omit any power losses in our fiber set-up, which simply scale the effective power $P$. Experimentally, one can simply measure the reflected power at quadrature $P_{\mathrm{DC}}$ to determine the correct scaling.

\subsection{Light to voltage conversion}
\label{sssec:shot}
\subsection*{Photodiode}
A photo detector is used to convert the reflected optical signal into a photo current, via the photodiode responsitivity  $r_{\mathrm{PD}}$ ($(r_{\mathrm{PD}}) = $ ampere/watt = A/W). The current response for the Airy function model gives
\begin{equation}
\Delta i_{R} =  r_{\mathrm{PD}} P \frac{\lambda_{0}}{\bar{z}_m} \left( \frac{\mathrm{d} R}{\mathrm{d}\lambda} \right) \Delta z_m  .
\label{eq:currentairy}
\end{equation}

\subsection*{Shot noise}

The DC power on the photodiode generates a shot noise, which can be modeled as an effective white photodiode current noise $\widetilde{i}_n$. This noise depends only on the DC input power and the photodiode responsitivity.
\begin{equation}
\widetilde{i}_n = \sqrt{2 q_e P_{\mathrm{DC}}   r_{\mathrm{PD}}        } 
\label{eq:shotnoise_i}
\end{equation}
The shot noise induced displacement noise depends on the ratio of this current noise to $\Delta i_{R} /z_m $. For the sinusoidal model this gives
\begin{equation}
\widetilde{z}_n = \frac{\widetilde{i}_n \cdot z_m}{\Delta i_{R}} = \sqrt{ \frac{ q_e \lambda_q^2}{r_{\textrm{PD}} \kappa^2 8 \pi^2 P_{\textrm{DC}}       } }.
\label{eq:shotnoise_z_sine}
\end{equation}
For the general cavity response this corresponds to
\begin{equation}
\widetilde{z}_n =  \sqrt{\frac{2 q_e R (z_m, \lambda_0) }{r_{\textrm{PD} }P} } \frac{\bar{z}_m}{   \left( \frac{\mathrm{d} R}{\mathrm{d} \lambda} \right) \lambda_0}.
\label{eq:shotnoise_z}
\end{equation}

\subsection*{Trans-impedance amplifier}

The conversion into readout voltage is done using a trans-impedance amplifier (TIA), which is often combined with the photodiode in a photo receiver. At this point, we assume that the TIA has either sufficient bandwidth to minimize any frequency dependent scaling effects, or that the back-end corrects for this. The TIA is then simply characterized by a gain $\mathcal{R}_\mathrm{TIA}$ that determines the current-to-voltage ratio.

The photo receiver elements can introduce additional electronic noise that can spoil the measurement performance. Care should be taken to reduce any such influence to negligible levels. A simple method to do this is to calculate the equivalent input current noise of all contributions in the photo receiver \cite{Cervantes2011} and to compare them to the expected shot noise level (see Equation \ref{eq:shotnoise_i}). 

\section{Error budget}

The overall error budget for the acceleration readout performance is compiled for the parameters listed in Table \ref{tbl:assumedparams}. 

\begin{table}
\caption{\label{tbl:assumedparams}List of parameter values used for calculating the error and accuracy limits.}
\begin{tabular}{ccc}
			\hline 
			parameter & value & unit\\
			\hline 
			$\omega_0$ & $2\pi\times 10646.56$ & Hz\\
			m & 25 & mg\\
			$r_{\textrm{PD}}$ & 0.95 & A/W\\
			$\lambda_0$ & 1558.5 & nm\\
			$\kappa$ & 0.97 & -\\
			$\bar{z}_m$ & 42 & $\upmu$m\\
			$T$ & 300 & kelvin (K)\\
			g & 9.81 & m/s$^2$\\
			$k_B$ &$ 1.3806488\times 10^{-23}$ & m$^2$\,kg\,s$^{-2}$\,K$^{-1}$\\
			$q_e$ & $1.602\times 10^{-19}$ & C\\
			$c$ & 299792458 & m/s\\
			\hline 
		\end{tabular}
\end{table}
The parameters are either chosen by design or they represent experimentally reproducible values. For the major error contributions we can now derive the necessary values of the other readout parameters to achieve a sensitivity of $10\,\upmu$g/\whz. The results are summarized in Table \ref{tbl:noise}. Thermal noise is completely negligible. Shot noise and RIN are broadband and can both dominate the high frequency behavior. The influence of RIN can, however, be easily reduced by implementing an amplitude stabilization. Laser frequency noise is expected to dominate at low frequencies for free running lasers, as observed. For a well-stabilized laser source the low frequency performance could at some point also be dominated by parasitic displacement noise, induced, for example, by thermal fluctuations that drive the non-zero coefficient of thermal-expansion of the device.
\begin{table}
\caption{\label{tbl:noise}Relevant noise sources and boundary conditions necessary to achieve a sensitivity of better than $10\,\upmu$g/\whz.}
\begin{tabular}{ccccc}
\hline 
\hspace{1.5cm}error & & & \multicolumn{2}{c}{value for reaching $\widetilde{a}$}\\
source  & symbol & & parameter & $< 10\,\upmu$g/\whz \\
\hline 
thermal & $\widetilde{a}_{\mathrm{th}}$ & & $Q$ & $>0.005$ \\
displacement & $\widetilde{a}_{\mathrm{d}}$ & & $\widetilde{z}_{\mathrm{d}}$ & $<22$\,fm/\whz\\
frequency & $\widetilde{a}_f$ & & $\widetilde{f}_n$ & $<101$\,kHz/\whz \\
amplitude & $\widetilde{a}_P$ & & RIN & $< 1.7 \times 10^{-7}$/\whz \\
shot & $\widetilde{a}_n$ & & $P_{\textrm{DC}}$ & $>11.3$\,$\upmu$W \\
\hline 
\end{tabular}
\end{table}

\section{Self calibration \& accuracy}
\label{sec_accuracy}
\begin{figure}
	\centering
	\includegraphics[width=8cm]{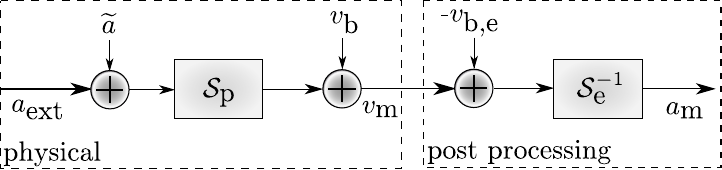}
	\caption{\label{fig:accuracymodel} Model of the metrology chain from external to measured acceleration that is used to determine the accelerometer performance.}
\end{figure}
We evaluate the performance of our accelerometer and its self calibration based on the model shown in Figure \ref{fig:accuracymodel}. Any external acceleration experienced by our device is contaminated by some measurement noise $\widetilde{a}$, which we discussed in the previous section, and is converted into a voltage with the actual physical scale factor, or, to be more precise, with the scale operator $\mathcal{S_{\textrm{p}}}$. A constant bias $v_{\textrm{b}}$ is added to form the total measured output voltage $v_{\textrm{m}}$. During post processing we subtract an estimated voltage bias $v_{\textrm{b,e}}$ and we convert into measured acceleration $a_{\textrm{m}}$ by applying the inverse of the estimated scale operator $\mathcal{S_{\textrm{e}}}^{-1}$. In this model we neglect any influence of scale factor inaccuracies on the determined noise /error levels. We can now write down the measured acceleration,
\begin{equation}
a_{\textrm{m}} =  \mathcal{S_{\textrm{e}}}^{-1} ( v_{\textrm{b}}- v_{\textrm{b,e}}  + \mathcal{S_{\textrm{p}}} ( a_{\textrm{ext}} +\widetilde{a}  )  ) 
\label{eq:accmodel1}
\end{equation}
We can write down the full analytical expression for the scale operator, based on Equations \ref{eq:funcaccdisp} and \ref{eq:airyfunceff},
\begin{equation}
\mathcal{S_{\textrm{e}}}^{-1} =  - \frac{\bar{z}_m}{\lambda_0} \left( \frac{\mathrm{d} v}{\mathrm{d} \lambda} \right)^{-1} \left( \frac{\mathrm{d}}{\mathrm{d} t^2} + \frac{\omega_0}{Q} \frac{\mathrm{d}}{\mathrm{d} t} + \omega_0^2 \right) .
\label{eq:accmodel2}
\end{equation}
Assuming we can subtract DC biases well and neglecting the influence of the scale factor inaccuracies on the noise/error levels, we simplify Equation \ref{eq:accmodel1} to
\begin{equation}
a_{\textrm{m}} =  \mathcal{S_{\textrm{e}}}^{-1}  \mathcal{S_{\textrm{p}}}  a_{\textrm{ext}} + \widetilde{a} .
\label{eq:accmodel3}
\end{equation}
If we now compute the effective error of our acceleration measurement $(a_{\textrm{err}} = a_{\textrm{m}}  -a_{\textrm{ext}} ) $ we get
\begin{equation}
a_{\textrm{err}} = a_{\textrm{ext}} \cdot ( \mathcal{S_{\textrm{e}}}^{-1}  \mathcal{S_{\textrm{p}}} -1) + \widetilde{a} .
\label{eq:accmodel4}
\end{equation}
We call $\mathcal{S_{\textrm{e}}}^{-1}  \mathcal{S_{\textrm{p}}} -1$ the scale factor accuracy $\mathcal{S}_{\sigma}$ operator, and, together with the noise and the dynamics of the external acceleration, they determine the total readout error.
\begin{equation}
a_{\textrm{err}} = a_{\textrm{ext}} \cdot \mathcal{S}_{\sigma} + \widetilde{a} .
\label{eq:accmodel5}
\end{equation}

\subsection{Scale factor accuracy}
\label{ss:scaleacc}

\begin{table}
\caption{\label{tbl:scale_accu} Overview of the components that limit the fundamental scale-factor accuracy}
\begin{tabular}{ccc}
\hline 
transducer  & mech. oscillator & interferometer\\
\hline 
parameter  & $\omega_0$, $ Q$ & {$\frac{\mathrm{d} v}{\mathrm{d} \lambda}$}, $\lambda_0$, $z_m$  \\
\hline 
SI reference & $f_{\rm ref}$ & {$\lambda_0$} \\
\hline 
simple model & {(viscously) damped } & two-beam interference \\
& {harmonic oscillator} &       (sinusoidal model)                 \\
& Equation \ref{eq:funcaccdisp} &     Equation \ref{eq:sine}  \\
\hline 
extensions & {higher order modes } & multiple reflections \\
& {non-viscous damping } & fiber coupling efficiency \\
& {non-linearity } & {non-linearity }  \\
\hline 
stability  & thermal expansion & thermal expansion \\
influence  & clamping          & intensity noise \\
\hline 
\end{tabular}
\end{table}

To discuss the estimated scale factor accuracy achievable with our device we determine the combined standard uncertainty $u_{{c}}$ for $\mathcal{S}_{\sigma}$. 
We summarize the relevant effects, models and fundamental parameters in Table \ref{tbl:scale_accu}, split by the two characterization steps that we use to estimate $\mathcal{S_{\textrm{e}}}^{-1}$. For a given model of the behavior of our system we can calculate the scale factor uncertainty based on measured, or estimated uncertainties of the relevant parameters. For a number of N parameters $x$ with relative uncertainties $u_x$ and estimated values $x_e$, given as $x_e = x(1+\varepsilon_x)$, applied with $\mathcal{S_{\textrm{e}}}^{-1}$, we estimate the combined standard uncertainty as
\begin{equation}
u_{c} = \sqrt{ \sum\limits_{x=1}^{N} \left(\frac{\partial \mathcal{S}_{\sigma} }{\partial  \varepsilon_x}  u_x \right)^2 }       .
\label{eq:scalefac}
\end{equation}
To give an example, we evaluate the coupling of $\omega_0$ into the standard uncertainty for low frequency signals. For this case  $\mathcal{S}_{\sigma}$ simplifies to 
\begin{equation}
\mathcal{S}_{\sigma} +1 = \frac{\omega_{0,e}^2}{\omega_{0}^2}  =
(1 + \varepsilon_{\omega_{0}})^2 = 1 + 2\varepsilon_{\omega_{0}} +\varepsilon_{\omega_{0}}^2.
\label{eq:scalefac2}
\end{equation}
Computing the derivative and omitting terms of the order $\mathcal{O}^2$ we determine a scaling factor of
\begin{equation}
\frac{\partial  \mathcal{S}_{\sigma} }{\partial  \varepsilon_{\omega_{0}}} \approx 2
\label{eq:scalefac3ex}
\end{equation}
We have summarized the uncertainties of the relevant parameters and their coupling factors in Table \ref{tbl:scalecalc}. For the mechanical oscillator part we include two cases, at low frequencies and at resonance. For the interferometer we include the uncertainties for the here-presented low-finesse readout and, for comparison, the values for the high-finesse readout from the earlier study \cite{Guzman2014}. In the following we discuss the individual contributions.

\subsection*{Mechanical oscillator}
Fundamentally, the scale factor is limited by the variations of $\omega_0$ and $Q$, due to temperature changes, causing thermal expansion or material property changes, and due to changes in the clamping or mounting of the accelerometer, which also influence $Q$. From our earlier spectroscopy measurements we find that $\omega_0$ is stable within $1 \times 10^{-6}$ in our laboratory thermal environment and Q can be estimated with an uncertainty of about $200 \times 10^{-6}$. In future studies we will investigate the temperature influence further, to get a more detailed understanding of the thermal environment necessary to achieve even better scale factor accuracies. 

\subsection*{Interferometer}
The calibration of the optical metrology chain combines many of the earlier described transducer steps into a direct response of output voltage over laser wavelength, which is used to determine the effective ratio of voltage over displacement $\frac{\mathrm{d} v}{\mathrm{d} z}$. The fundamental limitation of this approach is given by the wavelength uncertainty. The currently used laser system provides a resolution of 0.01\,nm during a wavelength scan. From this we can estimate the wave length uncertainty limit to $u_{\lambda} =7 \times 10^{-6}$.
Changes of the cavity length due to thermal expansion or of the reflected power, due to intensity noise, influence the scale factor over time. Power stabilizations that reach levels of better than $10^{-6}$ over long time scales are feasible and can be used, if higher scale factor precisions are aimed for.

We have characterized the current experimental limits for our cavity readout characterization using the Airy function model described in Equation \ref{eq:airyfunc} and the resulting $\frac{\mathrm{d} v}{\mathrm{d} z}$ (see Equation \ref{eq:airyfunceff}). Within this model the uncertainties of four parameters are relevant: $u_{\lambda}$,  $u_{\bar{z}_m}$, $u_{R}$ and $u_{v_\gamma}$. The standard errors of our Airy function fit point to an uncertainty for ${\bar{z}_m}$ of better than $10 \times 10^{-6}$. Taking the fundamental limit of $u_{\lambda}$ for this determination into account, we estimate  $u{\bar{z}_m} \approx 12 \times 10^{-6}$. The reflectivity of our fiber ends is only poorly estimated by the fit to $u_{R} \approx 10\%$, but we can determine it better using additional reflection and transmission measurements to about $u_{R} \approx 0.5\%$. The coupling of reflection variations into the scale factor is complex and is given by the derivative of the Airy function \cite{Smith2009}, we estimate the uncertainty scaling to $\approx 0.003$. However, our dominating noise term arises from uncertainties in estimating ${v_\gamma}$, which, based on our fits, can be determined with an uncertainty of $u_{v_\gamma} \approx 1.5\times 10^{-3}$. We can exclude an influence of the wavelength dependent photodiode response, because we monitored the laser power during wavelength sweeps with the same type of diode. The wavelength dependent discrepancy between the interference amplitudes estimated by our fits (see Figure \ref{fig:laser_scan_model}) indicates that we are limited by diffraction related effects. Models that include wavelength dependent mode propagation, as well as corrections due to cavity misalignments \cite{Wilkinson2011} can be applied in future studies to improve the understanding and accuracy of the reflection response.

With the current models, at low frequencies and for the low-finesse cavity we can calculate   
\begin{equation}
u_{c}  = \\ \sqrt{ (2u_{\omega_0})^2 +  
	(u_{\lambda})^2 + 
	(u_{\bar{z}_m})^2 + 
	(0.003u_{R})^2 + 
		(u_{v_\gamma})^2  }       .
\label{eq:scalefaccalv}
\end{equation}
Table \ref{tbl:scalecalc} shows the resulting combined standard uncertainties. We also include the corresponding values achieved for the high-finesse readout. 
With the current low-finesse device we reach uncertainties of $0.25\%$, which is similar to levels achieved at NMI calibration facilities. The characterization of the optical response is currently limiting; applying more complex cavity models \cite{Wilkinson2011} can improve this in future studies. The uncertainties achieved with the high-finesse system are lower by almost an order of magnitude, however, they are still limited by the optical response characterization. One should note that the high-finesse cavity in Ref. \cite{Guzman2014} was also mode matched, using a curved mirror to form a hemispherical resonator, potentially reducing the influence of beam propagation effects neglected in our Airy function analysis.

\begin{table}
\caption{\label{tbl:scalecalc} List of the individual relative uncertainties and their scaling, together with the derived combined standard uncertainties for signals at low frequencies and at oscillator resonance. 
}
\begin{tabular}{c|ccccc} 
\hline\hline 
uncertainty  &  \multicolumn{4}{c}{  harmonic oscillator}\\
  &  \multicolumn{2}{c}{  $\omega \ll \omega_0$} & \multicolumn{2}{c}{  $\omega = \omega_0$} \\ 
& value  & $ \frac{\partial  \mathcal{S}_{\sigma} }{\partial  \varepsilon_x}$ & value & $ \frac{\partial  \mathcal{S}_{\sigma} }{\partial  \varepsilon_x}$  \\
\hline
$u_{\omega_0}$ &  $0.1 \times 10^{-6}$ & 2 & $0.1  \times 10^{-6}$ & 1 \\
$u_{Q}$ &   $<0.2 \times 10^{-3}$ & 0 & $<0.2 \times 10^{-3}$ & 1 \\
\hline\hline 
 &    &  &  &  \\
\hline\hline 
 uncertainty  & \multicolumn{4}{c}{Fabry-P\'erot interferometer} \\
 
&  \multicolumn{2}{c}{R = 4\,$\%$} & \multicolumn{2}{c}{$\mathcal{F} = 1600$} \\ 
& value  & $ \frac{\partial  \mathcal{S}_{\sigma} }{\partial  \varepsilon_x}$  &   value & $ \frac{\partial \mathcal{S}_{\sigma} }{\partial \varepsilon_x}$ \\

$u_{\lambda}$ &    $ 7\times 10^{-6}$ &  1 & $ 7\times 10^{-6}$ &  1 & \\
$u_{\bar{z}_m}$ &   $ 5\times 10^{-6}$ & 1 & $ 7\times 10^{-6}$& 1 & \\
$u_{R}$ &   $ 5\times 10^{-3}$  & 0.003 & $ 78.4\times 10^{-6}$ & $1.33$ & \\
$u_{v_{\gamma}}$ &  $ 1.5\times 10^{-3}$ & 1 &  $ 309\times 10^{-6}$ & 1 & \\
\hline\hline 
&    &  &  &  \\
\hline\hline 
$u_{c} $ & \multicolumn{2}{c}{R = 4\,$\%$} &  \multicolumn{2}{c}{$\mathcal{F} = 1600$}\\
$\omega \ll \omega_0 $ & \multicolumn{2}{c}{$ 1.50\times 10^{-3}$} & \multicolumn{2}{c}{$315 \times 10^{-6}$}  \\
$\omega = \omega_0 $ & \multicolumn{2}{c}{$ 1.51\times 10^{-3}$} & \multicolumn{2}{c}{$369 \times 10^{-6}$} \\
\hline\hline 
\end{tabular}
\end{table}

\subsection{Bias stability}
\label{ss:biasstab}

For a given model of the dominating noise sources one can compute a corresponding Allan deviation behavior \cite{IEEEStd1293_2011,Woodman2007,El-Sheimy2008,Ferre-Pikal1999,Quinchia2013}. By comparing the two most dominant noise types, white noise and 1/f noise, one can compute the Allan deviation minimum and derive from that the expected bias stability.

A white noise of $\widetilde{a}_w = a_w \cdot 1/$\whz induces an Allan deviation slope of
\begin{equation}
\sigma_w(\tau) = \frac{a_w}{\sqrt{2\tau}}.
\label{eq:whiteallan}
\end{equation}
A 1/f noise of $\widetilde{a}_{1/f} = a_{1/f}/f \cdot 1/$\whz induces an Allan deviation slope of
\begin{equation}
\sigma_{1/f} (\tau) = 2\pi\frac{a_{1/f} }{\sqrt{6}} \sqrt{\tau}.
\label{eq:1_fallan}
\end{equation}
For the minimum value, which is often denoted as the bias stability, we can now calculate the optimal integration time,
\begin{equation}
\tau_{bs} = \frac{\sqrt{3}}{2\pi} \frac{a_{w} }{a_{1/f}} .
\label{eq:bstau}
\end{equation}
This leads to an estimate of the bias stability of
\begin{equation}
\sigma_{bs} \approx 2\sigma_w(\tau_{bs}) = \sqrt{ \frac{{4\pi}}{\sqrt{3}}  a_{w} a_{1/f} }.
\label{eq:bstab}
\end{equation}
Assuming a white acceleration noise floor of $\widetilde{a}_{w} = 8\,\upmu$g/\whz and a 1/f noise floor of $\widetilde{a}_{w} = 0.9\,$mg$/f / \sqrt{\textrm{Hz}}$ the integration minimum can be found at $\tau_{bs} \approx 2.5\,$ms and the corresponding bias stability is on the order of 0.2\,mg, which is consistent with our data presented in Figure \ref{fig:allan_dev_acc_g}. Using the presented formulas we can now extrapolate the 1/f laser frequency stability $f_{1/f}$ that is required to achieve a certain bias stability with our model. This is calculated as
\begin{equation}
f_{1/f} = \frac{c}{\lambda_0 z_m \omega_0^2}  \frac{\sigma_{bs}^2}{a_{w}} \frac{\sqrt{3}}{4\pi} .
\label{eq:eslasstab}
\end{equation}
With our cavity parameters and the earlier assumed white noise floor we require a 1/f laser frequency noise of less than $f_{1/f}/f/\sqrt{\textrm{Hz}} \approx 1700\,$Hz/f/\whz to achieve a bias stability of $1\,\upmu$g at integration times of 1.3 seconds.

\section{Summary and conclusion}

We have presented an accelerometer device that is a promising candidate for future applications requiring high bandwidth and in-situ self-calibration. We have conducted a detailed analysis of the device physics and derived the effects that influence noise, accuracy and stability. We achieve a bias stability of better than $0.3\times10^{-3}\,$g and an estimated scale factor uncertainty of $0.15\%$. For an input signal of 2\,g we can estimate a total acceleration measurement error on the order of 3.3\,mg, dominated by systematic effects in the characterization of the cavity response.

The bias stability of our device is currently limited by laser frequency noise and non-stationary thermal and acoustic noise coupling into the measurement through parasitic beams in the fiber set-up. At frequencies above 1\,kHz our readout noise is dominated by shot noise. The accuracy of our device is limited by the characterization of the reflectivity response of the Fabry-P\'{e}rot cavity.

The limits of the achievable accuracies for scale factor and bias stability have been discussed, both for low-finesse and for high-finesse readout. While the ultimate limits of self-calibration to SI standards are rather well understood based on the applied simple behavior models, targeted studies will be conducted in the future to measure the experimental variations of the scale factor over long time scales and its dependency on temperature. Applying a cavity model that takes diffraction effects and misalignments into account may further improve the self-calibration of the cavity response. Further experiments are also required to test the limits of the simple behavior models by, for example, performing detailed measurements of the harmonic oscillator response function. 

\section*{Disclaimer}
Certain commercial equipment, instruments, or materials are identified in this paper in order to specify the test and measurement procedure adequately. Such identification is not intended to imply recommendation or endorsement by the National Institute of Standards and Technology, nor is it intended to imply that the materials or equipment identified are necessarily the best available for the purpose.

\section*{Acknowledgements}
The authors would like to thank Aaron Kirchhoff for his help with the device assembly and vacuum sealing. We would also like to thank Jason Gorman, Thomas LeBrun, and Robert Lutwak for useful discussion. This work was supported by DARPA and ARO under grant W911NF-14-1-0681.

\end{document}